\documentclass[conference]{IEEEtran}
\IEEEoverridecommandlockouts
\usepackage{cite}
\usepackage{amsmath,amssymb,amsfonts}
\usepackage{algorithmic}
\usepackage{graphicx}
\usepackage{textcomp}
\usepackage[dvipsnames]{xcolor}
\def\BibTeX{{\rm B\kern-.05em{\sc i\kern-.025em b}\kern-.08em
    T\kern-.1667em\lower.7ex\hbox{E}\kern-.125emX}}

\usepackage{tabularx}
\usepackage{listings}
\usepackage{tikz}
\usepackage{amsmath}
\usepackage{booktabs}

\usetikzlibrary{arrows.meta, positioning, fit}

\usepackage{hyperref}

\usepackage{multirow}
\usepackage{longtable}
\usepackage{enumitem}
\usepackage{array}

\usepackage{blindtext}
\usepackage[most]{tcolorbox} 
\definecolor{block-gray}{gray}{0.95}

\newtcolorbox{zitat}[2][]{%
    colback=block-gray,
    grow to right by=-5mm,
    grow to left by=-5mm, 
    boxrule=0pt,
    boxsep=0pt,
    breakable,
    enhanced jigsaw,
    borderline west={4pt}{0pt}{gray},
    title={#2\par},
    colbacktitle={block-gray},
    coltitle={black},
    fonttitle={\normalsize\bfseries},
    attach title to upper={},
    #1,
}

\definecolor{darkblue}{rgb}{0.0, 0.0, 0.55}
\usepackage[strict]{changepage}

\usepackage{framed}

\definecolor{formalshade}{rgb}{0.95,0.95,1}

\newenvironment{formal}{%
  \MakeFramed{\advance\hsize-\width\FrameRestore}%
  \noindent\hspace{-4.55pt}% disable indenting first paragraph
  \begin{adjustwidth}{}{7pt}%
  \vspace{2pt}\vspace{2pt}%
}
{%
  \vspace{2pt}\end{adjustwidth}\endMakeFramed%
}

\usepackage{graphicx}

\begin{document}

\title{Transparency in App Analytics: \\Analyzing the Collection of User Interaction Data
\thanks{*This paper has been accepted at the 20th Annual International Conference on Privacy, Security \& Trust (PST2023)}
}

\author{
    \IEEEauthorblockN{Feiyang Tang}
    \IEEEauthorblockA{\textit{Norwegian Computing Center}\\
    Oslo, Norway}
    \and
    \IEEEauthorblockN{Bjarte M. \O stvold}
    \IEEEauthorblockA{\textit{Norwegian Computing Center}\\
    Oslo, Norway}
    }

\maketitle

\begin{abstract}
The rise of mobile apps has brought greater convenience and many options for users. However, many apps use analytics services to collect a wide range of user interaction data, with privacy policies often failing to reveal the types of interaction data collected or the extent of the data collection practices. This lack of transparency potentially breaches data protection laws and also undermines user trust. 
We conducted an analysis of the top 20 analytic libraries for Android apps to identify common practices of interaction data collection and used this information to develop a standardized \textit{collection claim} template for summarizing an app's data collection practices wrt. user interaction data. 
We selected the top 100 apps from popular categories on Google Play and used automatic static analysis to extract \textit{collection evidence} from their data collection implementations. Our analysis found that a significant majority of these apps actively collected interaction data from UI types such as View (89\%), Button (76\%), and Textfield (63\%), highlighting the pervasiveness of user interaction data collection.
By comparing the collection evidence to the claims derived from privacy policy analysis, we manually fact-checked the completeness and accuracy of these claims for the top 10 apps. We found that, except for one app, they all failed to declare all types of interaction data they collect and did not specify some of the collection techniques used.
\end{abstract}


%-------------------------------------------------------------------------------
\section{Introduction}\label{Sec:intro}

The rapid rise of mobile apps has revolutionized how we interact with technology, providing developers with a treasure trove of user interaction data through analytics services such as AppsFlyer \footnote{\url{https://www.appsflyer.com/}}, Flurry \footnote{\url{https://www.flurry.com/}}, and Firebase Analytics \footnote{\url{https://firebase.google.com/docs/analytics}}. This data, encompassing user actions like button taps, page scrolls, and video views, is invaluable for enhancing app functionality and user experience. However, the vague terminology often used in privacy policies, such as ``user's interaction with the service'', raises concerns about transparency. The lack of specificity leaves users uncertain about the extent and nature of the data being collected and its usage, potentially leading to mistrust and diminished app usage.

Transparency in data collection is a crucial factor influencing user trust \cite{Cysneiros2009AnIA}. It empowers users to make informed decisions about the data they share and its intended usage \cite{morey2015customer}. %However, data collected without clarity can be exploited for various purposes like targeted advertising and user profiling, thus undermining user trust.

An example of this ambiguity can be found in the Yr app, Norway's most popular weather app developed by the Norwegian Broadcasting Corporation (NRK). Despite collecting user interaction data to understand commonly used features, the app's privacy policy\footnote{\url{https://hjelp.yr.no/hc/en-us/articles/360003337614-Privacy-policy}} is vague regarding the collection of such data, as quoted below in the blue box. Our examination of NRK's privacy policy revealed no explicit information about Yr's data collection practices, leading to concerns about user trust in both the app and NRK.

\begin{formal}
\textbf{Analyze tools}
    \\
    ``We use different tools to track the use on our app and website. This information gives us valuable information such as most popular pages and on what times Yr is being used the most. No information that can identify persons are available for Yr.''
\end{formal}

Our examination of NRK's privacy policy\footnote{\url{https://info-nrk-no.translate.goog/personvernerklaering/?_x_tr_sl=no&_x_tr_tl=en&_x_tr_hl=en-US&_x_tr_pto=wapp}} revealed no specific information regarding Yr's data collection practices. The policy mainly focuses on NRK's news services and their ``interaction with the services'' collection practices. This obscurity concerning Yr app's data collection practices raises concerns, as it might undermine user trust in both the app and NRK as a whole.

\begin{figure*}[htb!]
\centering
    \begin{tikzpicture}
    \begin{scope}
    [every node/.style = {
    minimum width={width("xxxxxxxxxxxxxxxxxx")+2pt},
    minimum height=1.2cm,
    shape=rectangle, double, rounded corners, draw, ->}]
    \node(A)                    {App Privacy Policy};
    \node(B)       [below=1.5cm of A,align=left] {Relevant \\Text Fragments};
    \node(C)       [below=1.5cm of B,align=left]  {Standardized \\Collection Claim};
    \node(D)       [right=1.5cm of A,align=left] {Decompiled \\Mobile App};
    \node(E)       [below=1.5cm of D,align=left] {Relevant Bytecode \\and UI Files};
    \node(F)       [below=1.5cm of E]  {Collection Evidence};
    \end{scope}
    
    \draw[->,thick] (A) -- (B) node[text width=2cm, right, midway] {Extract related text};
    \draw[->,thick] (B) -- (C) node[text width=2.5cm, right, midway] {Standardize the text fragments};
    \draw[->,thick] (D) -- (E) node[text width=3cm, right, midway] {Identify analytics library invocation};
    \draw[->,thick] (E) -- (F) node[text width=2.5cm, right, midway] {Link invocation with UI widget};

    \node[draw=none, rounded corners=0.5cm, fill opacity=0.1, fill=blue, inner sep=8pt,label={[darkblue!80]right:\textbf{Artifact}}, fit=(A)(D)] (M) {};
    \node[draw=none, rounded corners=0.5cm, fill opacity=0.1, fill=blue, inner sep=8pt,label={[darkblue!80]right:\textbf{Relevant information}}, fit=(B)(E)] (U) {};
    \node[draw=none, rounded corners=0.5cm, fill opacity=0.1, fill=blue, inner sep=8pt,label={[darkblue!80]right:\textbf{Standardized form}}, fit=(C)(F)] (Z) {};
    \end{tikzpicture}
\caption{Overview of the approach for analyzing collection claims and evidence in apps.}
\label{fig:approach}
\end{figure*}
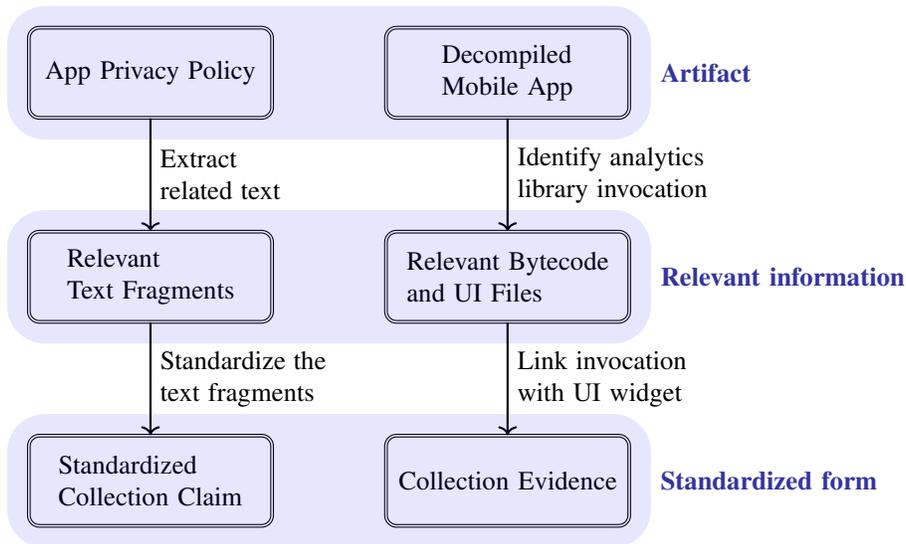

%Recent studies demonstrate that even seemingly innocuous user interaction data can disclose sensitive information about individuals. For example, emoji usage and pages visited can be employed to deduce a user's pool preference and political orientation, respectively~\cite{gadotti2022pool}. Furthermore, mobile biometric data related to keystrokes and touchscreen gestures can help estimate soft attributes like age, gender, and operating hand~\cite{7736910,jain2019gender}. These findings underscore the potential risks associated with collecting user interaction data, which can be utilized to infer sensitive information about individual users, leading to user profiling. Notably, such interaction data is not typically considered personal data and may remain unknown to users and unclear in privacy policy claims. This lack of transparency could jeopardize user trust in the app.
Recent research has shown that even seemingly harmless user interaction data can reveal sensitive information about individuals. For instance, data like emoji usage or pages visited can be used to infer a user's pool preference or political orientation \cite{gadotti2022pool}. Moreover, mobile biometric data related to keystrokes and touchscreen gestures can help estimate attributes like age, gender, and operating hand \cite{7736910,jain2019gender}. This underscores the potential risks associated with user interaction data collection, which, while not typically considered personal data, can be utilized to deduce sensitive information about individual users, leading to user profiling. The lack of transparency in these data collection practices could potentially erode user trust in the app.

Most current research on the privacy implications of analytic services has concentrated on determining whether personally identifiable information (PII) is being collected and transmitted to external analytics services~\cite{9284006,8660581}. Studies have also scrutinized log data to understand user behavior~\cite{dumais2014understanding}, and high-level analyses of user behavior data collection in mobile apps have been performed~\cite{verkasalo2010analysis}. It is essential to clarify that the interaction data discussed in this study is not part of traditionally defined PII or personal data, emphasizing the need for better transparency in data collection practices.

\subsection{Objective}
The aim of this paper is to address the issue of lack of transparency in the collection of user interaction data in mobile apps. To achieve this, we propose a standardized collection claim template that can be compared to collection evidence determined through static analysis.

\subsection{Research Questions}
To guide our investigation, we formulated the following research questions:

\begin{enumerate}[font={\bfseries},label=RQ\arabic*, ref=\arabic*]
   \item \label{rq:rq1} What are the common practices of user interaction data collection in mobile apps? 
   \item \label{rq:rq2} How are these practices reflected in apps' privacy policies? %We refer to these descriptions as \emph{collection claims}.
   \item \label{rq:rq3} What types of user interaction data do apps actually collect in their implementations, and how is this data collected? %We refer to this as the \emph{collection evidence}.
   \item \label{rq:rq4} To what extent do the collection claims in privacy policies align with the actual data collection practices as observed in their implementations?
   %\item \label{rq:rq4} To what extent do the collection claims align with the actual data collection practices as reflected in the collection evidence? 
\end{enumerate}

\subsection{Contributions}
In this paper, we make several contributions towards understanding and promoting transparency in user interaction data collection practices:

\begin{itemize}
    %\item We propose a \textit{standardized collection claim template} for user interaction data that reflects common phrasing and vocabulary derived from Android documentation and popular Android analytic libraries (Section~\ref{sec:claim}). This template can be used by app developers and researchers to provide a clearer description of data collection practices, facilitating better transparency and user trust.
    \item We propose a \textit{standardized collection claim template} for user interaction data. \textit{Collection claims}, in this context, refer to the descriptions of common practices of user interaction data collection in mobile apps, as stated in privacy policies. The template reflects common phrasing and vocabulary derived from Android documentation and popular Android analytic libraries (Section~\ref{sec:claim}). %This template can be utilized by app developers and researchers to provide a clearer and more standardized description of their data collection practices, thereby facilitating improved transparency and user trust.
    \item We present an automatic static analysis method to identify \textit{collection evidence} from Android apps, which involves analyzing data types, relevant code, and techniques of collection in layout files and bytecode (Section~\ref{Sec:analysis}). %This method can help researchers and app developers understand the types of user interaction data being collected and the collection metrics employed.
    \item We provide an overview of user interaction data collection practices in the top 100 popular apps on Google Play across the top 10 categories (Section~\ref{sec:finding}). Our analysis reveals common patterns and offers useful statistics for both app developers and users to better understand the current state of data collection practices in mobile apps.
    \item We conduct \textit{fact-checking} by manually comparing privacy policy collection claims against the actual collection evidence found in the ten most popular apps from the top ten categories (Section~\ref{sec:finding}). Our findings reveal that none of these apps were completely accurate and complete in their collection claims, highlighting the importance of our proposed approach in promoting transparency and trust in user interaction data collection practices.
\end{itemize}

We believe that our proposed method addresses the problem of lack of transparency by providing a standardized collection claim template for describing user interaction data collection practices. The template allows app developers to offer clearer and more accurate information about their data collection practices, which in turn can help users make informed decisions about app usage and data sharing. Our method also enables researchers and app developers to assess the alignment between stated data collection practices in privacy policies and the actual practices found in app implementations, facilitating better transparency and ultimately enhancing user trust.

Fig.~\ref{fig:approach} provides a high-level overview of our method, illustrating the process of acquiring artifacts (privacy policies and decompiled mobile apps), deriving knowledge through text analysis and static analysis (relevant text fragments and bytecode/UI files), and standardizing the information into collection claims and evidence. This systematic and transparent approach can contribute to promoting trust and fostering greater transparency in user interaction data collection practices in mobile apps.

\section{Motivation}
The transparency of mobile app data collection practices is a critical issue that stems from several significant factors, all of which play an essential role in the complex relationship between user trust and app adoption. Interaction data, a key asset for understanding user behavior, can raise serious concerns if it is collected non-transparently.

Transparency serves a dual purpose in this scenario. Firstly, it acts as an ethical commitment, assuring users that they are informed about their interaction data's collection and use. This principle not only respects user autonomy but also fosters an environment of openness and accountability. Secondly, transparency plays a pivotal role in building user trust, a significant factor influencing user satisfaction and continued app usage. When users understand and control their interaction data, their trust in the app increases, leading to more consistent engagement. Conversely, a lack of transparency can breed mistrust and privacy concerns, potentially causing user dissatisfaction or even app abandonment.

The correlation between transparency and user trust is well-documented in the academic world. Studies have consistently highlighted the positive relationship between increased transparency and elevated levels of user trust~\cite{fischer2016transparency}. Conversely, an absence of transparency can obstruct the success and widespread adoption of mobile apps~\cite{vorm2022integrating}. Thus, transparency is not merely about informing users, but is essential for facilitating informed decisions and granting consent.

Another critical aspect to consider is the rise of analytics services like Firebase Analytics and Flurry Analytics. These services provide developers with tools to gather data on user behavior, engagement, and preferences. However, they have raised data protection concerns as they often automatically collect user data, thereby creating privacy issues. Several countries, including France, Italy, Austria, Denmark, and Norway, have explicitly stated that the use of Google Analytics violates GDPR~\cite{SeveralE79:online}. Android apps can utilize these services either by directly invoking third-party APIs or by customizing their analytics service by extending these APIs. The first approach involves calling third-party API methods directly in activities to log user engagement events. The second approach allows developers to tailor data collection to their specific needs.

The importance of transparency is also acknowledged by mobile app developers, who have a vested interest in prioritizing it alongside user control in their data collection practices. Research supports this notion; for instance, Almuhimedi et al.~\cite{almuhimedi2015your}, for instance, discovered that many smartphone users are not fully aware of the data collected by their apps. Providing users with an app permission manager and sending notifications to increase their awareness of data collection can enable them to better manage their privacy. Moreover, users' concerns about data collection can negatively impact their perception of the app, potentially leading to its uninstallation~\cite{degirmenci2013mobile}. Hence, promoting transparency and user control can foster trust, resulting in improved user experiences, increased app engagement, and higher adoption rates.

In the following sections, we will explore the mutual benefits of transparency for both users and app developers and illustrate how our proposed method can address the current shortcomings in the transparency of interaction data collection practices.

\section{Standardizing Data Collection Claims}\label{sec:claim}
To address \textbf{RQ\ref{rq:rq1}}, we analyze the common practices of user interaction data collection in mobile apps, specifically the types of data collected and the techniques of collection. To achieve this, we conducted an analysis of the top 20 analytic libraries for Android apps.
To answer \textbf{RQ\ref{rq:rq2}}, we examine how these practices are reflected in the privacy policies of mobile apps. We refer to the descriptions of data collection practices in privacy policies as \emph{collection claims}, which we define as a single sentence in a standardized template using a restricted vocabulary to convey the essence of interaction data collection.

\subsection{Collection Vocabulary}\label{Sec:techvoc}
Our restricted collection vocabulary was developed by analyzing the Android system implementation documentation, as well as the APIs of the top 20 analytic services for Android apps listed on AppBrain~\cite{Androida43:online}.

\subsubsection{Terms for Types of User Interaction Data}
The user interface of an Android app collects a variety of data types, such as touch events, sensor data, and text input.  Based on a manual inspection of every single type of Android UI widget, we identified the following six types of interaction data and named them:

\begin{itemize}
    \item \emph{App presentation data}: This data arise from the consumption of content provided by the app. For example, the user plays a certain video for a period of time, spends minutes reading one specific page of the news. These interactions are often recorded by a logging system to keep track of the user's consumption habits.
    \item \emph{Binary data}: This data arise from discrete user actions, such as tapping on a button or icon, or selecting a checkbox. 
    \item \emph{Categorical data}: This data arise from a selection from a set of predefined options or categories, such as choosing a value from a dropdown menu, selecting a radio button, or rating a product. 
    \item \emph{User input data}: This data arise from user input through an on-screen keyboard or another input method, such as entering text or numbers into a form field, or using voice input to perform a search or command.
    \item \emph{Gesture data}: This data arise from gesture inputs and smooth and continuous movements of the user's finger on the screen, such as scrolling through a list, swiping left or right, pinching or zooming, or shaking the device. 
    \item \emph{Composite gestures data}: This data arise from a combination of multiple gestures, such as tapping and holding, double tag, or drag and drop.
\end{itemize}

\subsubsection{Terms for Collection Techniques}
We use the following terms to describe the techniques of user interaction data collection.
\begin{itemize}
    \item \emph{Frequency}: This technique involves logging the frequency of the occurrence of a particular interaction. For example, an app might log the number of times a user taps on a specific button or selects a certain option from a drop-down menu.
    \item \emph{Duration}: This technique involves tracking the time a user spends engaging in a particular interaction. For example, an app might log the amount of time a user spends watching a particular video or reading a specific article.
    \item \emph{Motion details}: This technique involves monitoring the specific details of a user's interaction, such as the speed, direction, or angle of their finger movements on the screen. This type of data can be collected for interactions such as scrolling, swiping, or dragging.
\end{itemize}

\begin{table*}[t]
\centering
\normalsize	
\resizebox{\textwidth}{!}{%
    \begin{minipage}{.5\textwidth}
      \caption{The top five most frequent terms used to describe user interaction data in the APP-350 corpus}
        \label{Tab:keyword1}
      \centering
        \begin{tabular}{lr}
            \toprule
            \textbf{Term} & \textbf{Count}\\ \midrule
            interact($\sim$ion,$\sim$ing) with service/app & 1,049 \\
            analytic(s) & 886 \\
            us($\sim$age, $\sim$ing) of service/app & 397 \\
            statistic(s) & 315 \\
            input(s) of user & 173 \\
            \bottomrule
        \end{tabular}
    \end{minipage}%
    \hspace{1cm}
    \begin{minipage}{.5\textwidth}
      \centering
        \caption{The top five most frequent verbs used to describe such collection in the APP-350 corpus}
        \label{Tab:keyword2}
        \begin{tabular}{lr}
            \toprule
            \textbf{Verb} & \textbf{Count}\\ \midrule
            collect & 1,386 \\
            track & 548 \\
            use & 202 \\
            log & 86 \\
            gather & 46 \\
            \bottomrule
        \end{tabular}
    \end{minipage} %
    }
\end{table*}

\subsection{From Policies to Standardized Collection Claims}\label{Sec:pp}
%We commence our analysis by scrutinizing privacy policies from publicly accessible mobile apps. For this purpose, we utilize the APP-350 Corpus~\cite{story2019natural}, which is composed of 350 mobile app privacy policies annotated with privacy practices. Given that APP-350's main focus lies in identifying sentences related to personally identifiable information (PII) from privacy policies, we employed their raw privacy policy HTML files for our examination.
In this section, we study privacy policies of publicly accessible mobile apps, aiming to identify and standardize collection claims related to user interaction data. Utilizing the APP-350 Corpus, a pre-trained language model, and manual checks, we extract and validate common terminologies employed in these policies. This comprehensive process allows us to establish a standardized vocabulary for user interaction data collection claims, providing a solid foundation for subsequent analysis.

\subsubsection{Identifying Relevant Policy Parts}
To distinguish sentences related to user interaction data collection in privacy policies, we adopt a simple pre-trained language model. This model sifts through HTML files of privacy policies and singles out sentences containing specific keywords and their synonyms. Our focus is to ensure that our privacy policy claim vocabulary aligns with the most common terminology used in the industry to describe user interaction data.

After processing the privacy policies, we conduct manual checks to eliminate any false positives. The end result is a selection of common phrases used in these policies to describe the collection of user interaction data. From the sentences identified, we isolate the most relevant verbs and nouns to form a list of keywords.

\paragraph*{Experimental Details and Validation}
In conducting our analysis, we use the APP-350 Corpus~\cite{story2019natural}, a set of 350 mobile app privacy policies that are annotated with privacy practices. Although the main focus of the APP-350 Corpus is on identifying sentences related to personally identifiable information (PII), we utilize the raw HTML files of the privacy policies for our examination.

The natural language processing is carried out using the spaCy~\cite{spacy2} library with the \verb|en_core_web_sm| model. This model, pre-trained on web text, which includes web forums, web pages, and Wikipedia, is capable of identifying named entities, parts of speech, dependency parsing, and more. We also employ the WordNet module from the Natural Language Toolkit (NLTK~\cite{bird2009natural}) to discover synonyms for the extracted keywords.

To authenticate the effectiveness of the model, we manually annotate 50 randomly selected privacy policies from the APP-350 dataset. This helps us identify sentences containing relevant information, the verbs used to describe data collection (e.g., ``collect'', ``track''), and the terms used to describe user interaction data (e.g., ``usage of the app'', ``interaction with the service'').

The model successfully recognized sentences related to user interaction data collection in 37 out of the 38 files that contained such sentences, using keywords such as interaction, usage, statistics, experience, and analytics. Identifying the verbs used to describe data collection was a more complex task, with a recall of 92\% but a precision of only 84\%
\footnote{Recall is calculated as TP / (TP + FN), while precision is calculated as TP / (TP + FP), where TP is true positives, FN is false negatives, FP is false positives, and TN is true negatives.}
due to the presence of similar verbs in sentences that were not related to the context.

Upon running the model on the 350 privacy policies, we identified 1,411 sentences. The relevant verbs and nouns from these sentences are shown in Table~\ref{Tab:keyword1} and Table~\ref{Tab:keyword2} and then compiled to form the list of keywords.

\subsubsection{Template for Standardized Collection Claims}
In privacy policies, it is common for apps to use convoluted language to describe how user data is collected. 
To make these collection claims in privacy policy easier to read and compare across different apps, we created a standardized template that utilized the most frequently used verb, ``collect'', and the most frequently used noun phrase, ``user interaction data''.
The resulting structure is as follows:

\begin{formal}
\textbf{Template for Standardized Collection Claims}
    \\
    We collect the following types of user interaction data: $\langle$\textit{types of data collected}$\rangle$, along with their $\langle$\textit{techniques of collection}$\rangle$.\footnotemark
\end{formal}
\footnotetext{Refer to the claim vocabularies in Section~\ref{Sec:techvoc}}

This standardized collection claim template can be combined with the collection evidence gathered through static analysis to check and the accuracy of privacy policy collection claims made by various apps. Also, the standardized language facilitates transparency and comparison between policies. 
We return to the subject of fact-checking collection claims in Section~\ref{sec:finding}.

\section{Data Collection Evidence}\label{Sec:analysis}
In this section, we analyze mobile apps to understand the types of user interaction data collected and the techniques employed (\textbf{RQ\ref{rq:rq3}}). We conduct static analysis of the Android application package (APK) to identify data collection methods (DCMs) and extract collection evidence, which highlights the gap between privacy policy claims and actual practices (\textbf{RQ\ref{rq:rq4}}).

Our analysis is divided into two parts. First, we identify DCMs from the top 20 Android analytic services and customized analytics services. Second, we extract collection evidence by focusing on invocations to analytics services, associated UI widgets, and the callbacks triggered by registered listeners.
%To understand the types of user interaction data collected by mobile apps and the techniques of collection (\textbf{RQ\ref{rq:rq3}}), we analyze the apps' implementations. We identified the data collection methods used in the app package (APK) and extracted relevant evidence using static analysis techniques. 
%We refer to this process as the \emph{collection evidence}. 
%To evaluate the extent of agreement between the actual data collection practices and the claims made in the privacy policy (\textbf{RQ\ref{rq:rq4}}), we conducted fact-checking by comparing the collection evidence extracted from the app with the collection claims made in the privacy policy.

\subsection{Identifying Data Collection Methods} 
Data collection methods (DCMs) are methods defined by analytics services, such as Firebase Analytics, that allow app developers to log user interaction data. 
DCMs provide a standardized way for app developers to collect user interaction data and track app usage in order to analyze and understand user behavior.

For example, the Firebase Analytics API provides the \texttt{logEvent()} method to log user events, such as button clicks or screen views.
Suppose we have a button \texttt{myButton} in the app's UI, and we want to track when the user clicks on it. 
We can do this using Firebase Analytics by adding the following code to the button's \texttt{OnClickListener}:

    \begin{lstlisting}[basicstyle=\ttfamily\scriptsize]
    myButton.setOnClickListener(new View.OnClickListener() {
        public void onClick(View v) {
            FirebaseAnalytics.getInstance(this).
                logEvent("button_click", null); }
    });
    \end{lstlisting}

Here \texttt{FirebaseAnalytics.getInstance(this)} returns an instance of the Firebase Analytics object, and \texttt{logEvent("button\_click", null)} collects the button click interaction data with the string \texttt{"button\_click"} to Firebase Analytics. 

To determine how Android apps use analytics services, we identified DCMs from the top 20 Android analytic services, cf.\ Section~\ref{Sec:techvoc}.
Matching the full signature of these methods in bytecode allows us to find direct invocations to analytics services.
However, some apps use customized analytics services to do a more fine-grained collection, such as collecting motion details and duration. To do this, the apps implement their own analytics classes by extending the analytic services.

To identify customized analytics, we use static analysis to identify the classes that invoke external DCMs. 
We then check whether these classes are invoked in any of the app's declared activities. 
If they are, we mark these classes as customized analytics services classes. 

\subsection{Extracting Collection Evidence}
Next, we extracted evidence of actual data collection from the APK. 
Specifically, we analyze three types of information: (1) invocations to analytics services that logged user interaction data collection, (2) associated UI widgets, and (3) the callbacks triggered by registered listeners on these UI widgets.

We utilized static analysis with FlowDroid~\cite{arzt2014flowdroid} to associate DCM invocations with callbacks, listeners, and activities in the bytecode. 
We then compared the layout IDs of the associated UI widgets defined in the layout XML files to identify the relevant collection data types and techniques. 

\begin{figure}[htbp]
\centering
    \begin{tikzpicture}
    \begin{scope}
    [every node/.style = {shape=rectangle, rounded corners, draw, align=center, ->}]
    \node(A)                    {App};
    \node(B)       [below=of A] {Layout files};
    \node(C)       [left=of B]  {Activity};
    \node(D)       [below=of B] {UI widget};
    \node(E)       [below=of C,left=of D] {Listener};
    \node(F)       [left=of E]  {Callback};
    \node(G)       [above=of F,left=of C] {DCM \\ Invocation};
    \end{scope}
    
    \draw  (A) to node [right, near end] {\tiny M} node [right, near start] {\tiny 1} (B);
    \draw  (A) to node [left, near end] {\tiny M} node [left, near start] {\tiny 1} (C);
    \draw  (B) to node [right, near end] {\tiny M} node [right, near start] {\tiny 1} (D);
    \draw  (C) to node [right, near end] {\tiny M} node [right, near start] {\tiny 1} (D);
    \draw  (C) to node [left, near end] {\tiny M} node [left, near start] {\tiny 1} (E);
    \draw  (C) to node [left, near end] {\tiny M} node [left, near start] {\tiny 1} (F);
    \draw  (C) to node [above, near end] {\tiny *} node [above, near start] {\tiny 1} (G);
    \draw  (F) to node [left, near end] {\tiny 1} node [left, near start] {\tiny *} (G);
    \draw  (E) to node [above, near end] {\tiny M} node [above, near start] {\tiny 1} (F);
    \draw  (E) to node [above, near end] {\tiny 1} node [above, near start] {\tiny 1} (D);

    \node[draw=none, rounded corners=0.3cm, fill opacity=0.1, fill=blue, inner sep=10pt,label={[blue!80]above:Techniques of collection}, fit=(C)(E)(F)(G)] (M) {};
    \node[draw=none, rounded corners=0.3cm, fill opacity=0.1, fill=red, inner sep=10pt,label={[red!80]below:User interaction data types}, fit=(B)(D)] (U) {};
    \end{tikzpicture}
\caption{Relationships between different parts of the extracted collection evidence in an Android app} 
\label{fig:relationship}
\end{figure}
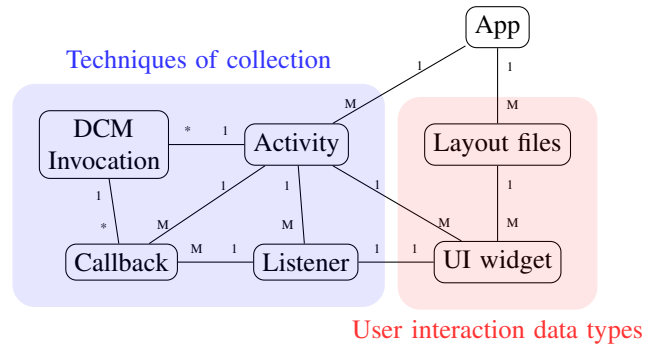

The relationships between different parts of the extracted collection evidence in an Android app are shown in Fig.~\ref{fig:relationship}.
The UI-related parts, such as layout files and defined UI widgets, provide information on the types of user interaction data (red section), while the bytecode provides details on the techniques of collection (blue section)\footnote{Note: The figure notation is as follows: 1-M means one-to-many, 1-$\ast$ means one-to-any (zero or more), and 1-1 means one-to-one.}.

We return to the Yr weather app, the example app from Section~\ref{Sec:intro}. Based on the collection evidence extracted from Yr's bytecode and layout files, we discovered that it collects detailed user interaction data using various types of UI widgets such as \texttt{SearchView} and \texttt{Textfield}. This data collection is linked to features such as changes in location, enabling forecast summary notifications, and opening the forecast graph. Building on this finding from static analysis, we propose the following more specific checked standardized collection claim:

\begin{formal}
    \textbf{Checked Standardized Collection Claim for Yr}
    \\
    We collect the following types of user interaction data: \textit{app presentation, binary and categorical interactions, and user input interactions}, along with their \textit{frequency}.
\end{formal}

\section{Findings}\label{sec:finding}

To address \textbf{RQ\ref{rq:rq2}}, we conduct a manual inspection of 1411 sentences that described user interaction data collection in all 350 privacy policies within the APP-350 corpus, as outlined in Section~\ref{Sec:pp}.

We examine whether the sentences in a privacy policy provide clear descriptions of the types of user interaction data collected and the techniques of collection.

%\begin{tcolorbox}
We find that only 37\% of the identified sentences contained clear statements on both the data types and techniques of collection, while 41\% only discussed the techniques of collection and 22\% mentioned only the data types.
%\end{tcolorbox}

Here are the relevant sentences from two policies in the corpus. DAMI\footnote{\url{https://play.google.com/store/apps/details?id=com.blappsta.damisch}} states: ``\textit{We may work with analytics companies to help us understand how the Applications are being used, such as the frequency and duration of usage.}'' Wish\footnote{\url{https://play.google.com/store/apps/details?id=com.contextlogic.wish}} states: ``\textit{We may collect different types of personal and other information based on how you interact with our products and services. Some examples include: Equipment, Performance, Websites Usage, Viewing and other Technical Information about your use of our network, services, products or websites}.''

DAMI's privacy policy only discloses the techniques of collection, such as the frequency and duration of usage, without clearly explaining which type of user interaction data is collected. In contrast, Wish's privacy policy does mention some specific types of data collected, such as equipment and performance data, but it is unclear about which techniques of collection are used. 

The majority of identified sentences discuss the techniques of collection rather than specific data types, suggesting that organizations use the tactic of avoiding or minimizing disclosures about the types of user interaction data they collect in order to collect more data than users are aware of or comfortable with. 

To investigate \textbf{RQ\ref{rq:rq3}}, we performed a static analysis on a sample of 100 free Android apps downloaded from the top 10 most popular categories on the German Google Play store\footnote{\url{https://play.google.com/store/apps?gl=DE}}, as identified by AppBrain.
In cases where the same app appeared in several categories, we moved to the next popular app in the second category to get a total of 100 distinct apps. 

\begin{table*}[htbp]
\caption{Statistics of the user interaction data collection for the top 100 Android apps.}
\label{Tab:appresults}
%\footnotesize
\centering
\resizebox{\linewidth}{!}{%
    \begin{tabularx}{\linewidth}{lllrr} \toprule
    \textbf{UI type (types of interaction data)} &
      \textbf{Top 2 techniques of collection} &
      \textbf{Top 3 app categories} &
      \textbf{Percent collected} &
      \textbf{Avg \# collected} \\ \midrule
    View (Presentation)               & Frequency (100\%), Duration (52\%) & Entertainment, Shopping, Travel & 89\% & 12 \\
    Button (Binary)                   & Frequency (94\%), Motion (8\%)     & Social, Utility, Gaming         & 76\% & 26 \\
    Textfield (Input)                 & Frequency (100\%), Duration (4\%)  & Social, Shopping, Utility       & 63\% & 5  \\
    Checkbox \& Spinner (Categorical) & Frequency (97\%), Motion (16\%)    & Shopping, Travel, Utility       & 32\% & 7  \\
    GestureDetector (Gesture)         & Motion (94\%), Duration (40\%)     & Gaming, Entertainment, Social   & 16\% & 38 \\ \bottomrule
    \end{tabularx}%
    }
\end{table*}

%\begin{small}
%\begin{tcolorbox}
Our analysis of the top 100 Android apps revealed that app developers placed a great deal of emphasis on understanding how frequently users interacted with different UI elements (which may correspond to different features or functionalities in the app), as frequency was the top techniques of collecting user interaction data across all UI types. We also found that the average number of interaction data collected varied significantly across different types of UI. It was also interesting to see that the high number of interaction data collected for the button UI type (also found in 76\% of the apps), indicated that understanding button usage was a particularly important metric for app developers.
%\end{tcolorbox}
%\end{small}

%Table~\ref{Tab:appresults} compares collection practices across various app categories. 
%Gaming apps collect a high percentage of Gesture data and send it to the analytics services later, likely due to the importance of tracking user finger motions for an optimal gaming experience. 
%Entertainment, Shopping, and Travel apps tend to collect a high percentage of View data, indicating the importance of visual design customization and its consequences for the revenue stream of the app's service.
%Social and Utility apps concentrate on Button and Textfield data, reflecting their dependence on user input to perform actions and provide information. 
%In summary, the result reveals differences in the app categories that prioritize understanding user interaction data depending on the UI type. 
%This suggests that app developers may be focusing their efforts on understanding certain aspects of user behavior, such as how users view content, interact with buttons, or use gestures, depending on the specific goals and requirements of their app. 
%Such insight could help app developers better tailor their data collection and analysis strategies to achieve their specific objectives.
Table~\ref{Tab:appresults} presents an overview of the user interaction data collection practices across various app categories, focusing on the top UI type for each type of interaction data. We have selected the most frequently occurring UI types from each category for this analysis, which are listed in the first column.

The second column indicates the top two techniques linked with each UI type. The percentages in parentheses, for instance, 100\% and 52\% for the ``View'' UI type, represent the proportion of apps that use a particular technique in tracking the UI type. For example, 100\% of apps tracked ``View'' interactions use the frequency technique, while 52\% also use the duration technique.

The ``Percent collected'' column indicates the proportion of the top 100 apps that collect data related to a specific UI type. For instance, ``View'' data is collected by 89\% of the analyzed apps.

Finally, the ``Average \# collected'' column represents the average number of distinct DCMs detected in each app associated with a particular UI type. For example, on average, 12 distinct DCMs were detected for ``View'' data collection across the analyzed apps.

Upon comparing these user interaction data collection practices with the declarations in privacy policies, we observe a larger mismatch in certain app categories. Gaming apps, despite their high prevalence of Gesture data collection to optimize user experience, often lack comprehensive disclosure of such practices in their privacy policies. Similarly, Entertainment, Shopping, and Travel apps extensively collect ``View'' data, but their policies rarely match the extent of this data collection, indicating a transparency gap in these visually-centric applications.

Social and Utility apps, which heavily rely on ``Button'' and ``TextField'' data, also demonstrate a significant disparity between their actual data collection practices and policy disclosures. These observations highlight that while app developers tailor their data collection strategies to their specific objectives and requirements, they often fail to mirror this granularity in their privacy policies.

This mismatch is consequential as it affects the transparency of these apps and the users' ability to make informed decisions. Hence, addressing these discrepancies becomes crucial, and our findings provide valuable insights for developers aiming to improve their privacy disclosures, ultimately fostering trust and success in the app ecosystem.

To address \textbf{RQ\ref{rq:rq4}}, we manually inspected the privacy policy claims of the most popular app in each of the 10 categories on Google Play. We generated our checked collection claims by analyzing the actual data collection practices of each app and comparing them to the privacy policy claims published by the app.
Our checked collection claims are made by combining the evidence gathered through static analysis and the proposed standardized claim template.
The results are fact-check collection claims presented in Table~\ref{Tab:factcheck}.

%\begin{tcolorbox}
Our findings uncovered inconsistencies between the claims made in privacy policies and the actual data types and techniques of collection used by popular apps on Google Play. Many apps do not fully disclose the types of data collected or the techniques of collection, often using vague language such as ``collecting user interactions to improve the service''. 
%\end{tcolorbox}

Notably, some apps that may be perceived as having questionable data collection practices, such as TikTok and Amazon Prime Video, actually provided more detailed information on the types of data collected and the techniques of collection used. TikTok and Duolingo even provided specific examples of their data collection practices.

\begin{table*}[htbp]
\caption{Fact-checked data collection claims w.r.t.\ evidence for the most popular app from each of the top 10 categories of Google Play. \\The {\color{red!60}red text} indicates types of user interaction data missing from the privacy policy/collection claims, while the {\color{blue!60}blue text} indicates undisclosed techniques of collection.}
\label{Tab:factcheck}
\small
\centering
\resizebox{\textwidth}{!}{%
\begin{tabular}{p{0.34\textwidth}@{\hskip 0.2in}p{0.65\textwidth}}\toprule
  \textbf{Checked Collection Claim} &
  \textbf{Related Text in the Published Privacy Policy} \\ \midrule
%TikTok &
  \textbf{[TikTok]} We collect the following types of user interaction data: app presentation, binary, {\color{red!60} categorical}, user input, gesture and composite gesture interactions, along with their frequency, duration and {\color{blue!60}motion details}. &
  \textbf{[TikTok]} We collect information about how you engage with the Platform, including information about the content you view, the duration and frequency of your use, your engagement with other users, your search history and your settings.\\
%SHEIN &
  \textbf{[SHEIN]} We collect the following types of user interaction data: app presentation, binary, categorical, {\color{red!60} user input interactions}, along with their {\color{blue!60} frequency and duration}. &
  \textbf{[SHEIN]} Data about how you engage with our Services, such as browsing, adding to your shopping cart, saving items, placing an order, and returns for market research, statistical analysis, and the display of personalized advertising based on your activity on our site and inferred interests; Collect your device information, and usage data on our website or app for fault analysis, troubleshooting, and system maintenance, as well as setting default options for you, such as language and currency. The display of information you choose to post on public areas of the Services, for example, a customer review. \\
%Booking.com &
  \textbf{[Booking.com]} We collect the following types of user interaction data: {\color{red!60}app presentation, binary, categorical, user input interactions}, along with their {\color{blue!60}frequency and duration}. &
  \textbf{[Booking.com]} We collect data that identifies the device, as well as data about your device-specific settings and characteristics, app crashes and other system activity. \\
%PayPal &
  \textbf{[PayPal]} We collect the following types of user interaction data: app presentation,  {\color{red!60}binary, categorical, user input interactions} along with their  {\color{blue!60}frequency}. &
  \textbf{[PayPal]} When you visit our Sites, use our Services, or visit a third-party website for which we provide online Services, we and our business partners and vendors may use cookies and other tracking technologies to recognize you as a User and to customize your online experiences, the Services you use, and other online content and advertising; measure the effectiveness of promotions and perform analytics; and to mitigate risk, prevent potential fraud, and promote trust and safety across our Sites and Services. \\
%Duolingo &
  \textbf{[Duolingo]} We collect the following types of user interaction data: app presentation, binary, categorical, user input, gesture interactions, along with their {\color{blue!60}frequency} and duration. &
  \textbf{[Duolingo]} We do record the following data: Patterns, Clicks, Mouse movements, Scrolling, Typing, Pages visited, Referrers, URL parameters, Session duration. \\
%Amazon Prime Videos &
  \textbf{[Amazon Prime Videos]} We collect the following types of user interaction data: app presentation, binary, categorical, user input,  {\color{red!60}gesture interactions}, along with their  {\color{blue!60}frequency, duration and motion details}. &
  \textbf{[Amazon Prime Videos]} We automatically collect and store certain types of information about your use of Amazon Services including your interaction with content and services available through Amazon Services. List of examples: search for products or services in our stores and download, stream, view, or use content on a device, or through a service or application on a device. \\
%Yazio &
  \textbf{[Yazio]} We collect the following types of user interaction data: binary and  {\color{red!60}user input} interactions, along with their  {\color{blue!60}frequency}. &
  \textbf{[Yazio]} The Firebase Analytics service helps to determine the interactions of App users by recording, for instance, the first time the App is opened, deinstallations, updates, system crashes and how often the App is used. The service also records and analyses certain user interests. \\
%Fasion Famous &
  \textbf{[Fasion Famous]} We collect the following types of user interaction data:  {\color{red!60}app presentation, binary, user input, gesture and composite gesture interactions}, along with their  {\color{blue!60}frequency, duration and motion details}. &
  \textbf{[Fasion Famous]} Information that may be collected automatically: Data and analytics about your use of our Services. Data we collect with cookies and similar technologies: Data about your use of our Services, such as game interaction and usage metrics. \\
%Picsart &
  \textbf{[Picsart]} We collect the following types of user interaction data:  {\color{red!60}app presentation, binary, gesture and composite gesture interactions}, along with their  {\color{blue!60}frequency, duration and motion details}. &
  \textbf{[Picsart]} Our servers passively keep an electronic record of your interactions with our services, which we call ``log data''. We collect and combine data about the devices you use to access Picsart, and data about your device usage and activity. \\
%Dezor &
  \textbf{[Dezor]} We collect the following types of user interaction data: app presentation, binary,  {\color{red!60}categorical, user input interactions}, along with their  {\color{blue!60}frequency}. &
  \textbf{[Dezor]} The information collected by log files include internet protocol (IP) addresses, browser type, Internet Service Provider (ISP), date and time stamp, referring/exit pages, and possibly the number of clicks. \\ \bottomrule
\end{tabular}%
}
\end{table*}

However, we found that some apps from less controversial categories, such as the photography editing app Picsart and the payment platform PayPal, used opaque language in their privacy policies, leaving a large gap between their claims and our findings. The most extreme example was Booking.com, which extensively collects user interactions within the app, yet discloses almost no information in its privacy policy. These findings highlight the need for clearer and more comprehensive disclosures in privacy policies, particularly for apps that collect sensitive user data.

\subsection{Threats to Validity}

Potential threats to the validity of our experiment may impact the interpretation of our findings. A primary limitation of our experiment is the number of apps we manually fact-checked for data collection practices. Due to the complexity of accommodating varying UI types and callbacks into our predefined six data types and three techniques of collection, we were only able to manually fact-check one app in each category, totaling ten apps. This sample size, though limited, may not encapsulate the full diversity of data collection practices across all apps.

Furthermore, measuring the recall of our analysis posed a considerable challenge, given the absence of a comprehensive ground truth detailing all interaction data collection practices in each app. Consequently, our findings may not wholly represent the full range of data collection practices.

\section{Related Work}
The related work can be categorized into three primary themes: (1) privacy policy analysis using NLP and policy compliance check, (2) static analysis for security and privacy in apps, and (3) analytics services analysis.

\subsection{Privacy Policy Analysis}
Numerous studies have focused on analyzing and improving privacy policies in mobile apps. Researchers have explored various NLP approaches to automatically process and understand privacy policy texts, as well as to assist users in comprehending these policies more effectively~\cite{10.1145/3180445.3180447, ramanath2014unsupervised, ravichander2021breaking}. However, these studies do not specifically address the issue of user interaction data collection, which is a significant gap that our research addresses. Tools like PrivacyFlash Pro~\cite{zimmeck2021privacyflash} and AutoCog~\cite{qu2014autocog} have been developed to audit privacy policy compliance by comparing disclosed policies with actual app behavior, but they primarily focus on personal data, not user interaction data. A recent study by Bardus et al.~\cite{bardus2022data} systematically mapped existing contact-tracing apps and evaluated the permissions required and their privacy policies, but it did not delve into the specifics of user interaction data collection.

\subsection{Static Analysis for Security and Privacy}
The static analysis approach has been used to enhance security and privacy in mobile apps. This involves analyzing app bytecode, identifying data leaks, and detecting privacy violations~\cite{avdiienko2015mining,enck2014taintdroid,zhang2020does}. Despite the progress in this field, there remains an underrepresentation of studies targeting user interaction data, a type of data often overlooked in privacy policies and their corresponding analyses. A novel system, LocationScope, was presented by Lu et al.~\cite{lu2023detecting} to detect and measure aggressive location harvesting in mobile apps at scale, but it did not specifically target user interaction data.

\subsection{Analytics Services Analysis}
Another line of research has concentrated on the role of analytics services in capturing user data, primarily focusing on PII. Alde~\cite{8660581}, for example, proposed a method employing both static and dynamic analysis to detect the key information gathered by analytics libraries, which are largely device-level data. PAMDroid~\cite{9284006} takes a similar approach, identifying personal data funneled into analytics services and treating it as a misconfiguration. The domain of user interaction data collection, however, remains relatively untouched in these studies. A recent study by Laperdrix et al.~\cite{laperdrix2022price} presented a privacy analysis of free and paid games in the Android ecosystem, but it did not specifically focus on user interaction data collection.

These studies have contributed to the understanding of privacy policies and data collection practices in mobile apps. However, there is a lack of research specifically on the practices of user interaction data collection and the transparency of related claims in privacy policies. Our work extends the scope of previous research by focusing on user interaction data collection practices and providing an analysis on comparing privacy policy disclosures with actual app behavior. This approach aims to enhance transparency and trust in the mobile app ecosystem, addressing the research gaps in the existing literature.

\section{Conclusion and Future Work}

In conclusion, our analysis of the top 100 apps uncovers the widespread collection of user interaction data, while the detailed examination of the top 10 apps reveals that privacy policies often inadequately disclose such practices. To address this lack of transparency, we introduced a standardized collection claim template that aids app developers in accurately detailing their data collection practices. This approach fosters informed decisions by users and enhances transparency by allowing assessments of alignment between declared and actual data collection practices for the manually analyzed apps.
Our findings lay the groundwork for improving data collection transparency in mobile apps and highlight the need for automating the policy-to-claims analysis. This insight could potentially guide future research and policy-making to foster a more secure and trustworthy app ecosystem.

Our approach has limitations that can be addressed in future research to improve the analysis of data collection practices. The current analysis only covers the top 20 analytics services and is confined to Android apps. Furthermore, the manual fact-checking of the top 10 apps relies on our interpretation of their policies. To overcome these limitations, machine learning models could be employed to automatically identify and categorize data collection methods (DCMs) within app code, reducing the need for manual analysis. This would involve training models to detect DCMs and categorizing them based on the data types and collection techniques they employ. Additionally, a more precise and fine-grained policy analysis could be developed to automatically extract interaction data types and collection techniques from privacy policies. By combining these advancements, we could create a fully automated approach to fact-check collection claims against the collection evidence, thereby increasing the efficiency and accuracy of analyzing data collection practices in mobile applications.

Another potential area for future work is the exploration of user studies to understand users' perceptions of interaction data collection practices and their impact on users' trust and app usage. Extending the analysis to include other platforms and analytics services could also contribute to a more holistic understanding of user interaction data collection practices across the mobile app ecosystem.

\section*{Acknowledgement}
We appreciate the valuable insights provided by Professor Staal Vinterbo. This work is part of the Privacy Matters (PriMa) project. The PriMa project has received funding from European Union’s Horizon 2020 research and innovation program under the Marie Skłodowska-Curie grant agreement No. 860315.

 \bibliographystyle{IEEEtran}
 \bibliography{ref}

\end{document}